

\overfullrule=0pt
\nopagenumbers
\magnification=1100
\voffset -1.2truecm  
\vsize=26truecm
\hsize=17.2truecm

\def\bls#1{\baselineskip=#1truept 
}

\def\title#1{\vbox to 2.5truecm{\parindent=0pt{#1}\vfil}  }
\def\authors#1{\vbox to 4.6truecm{\parindent=0pt{#1}\vfil}  }
\def\header#1{
$ $

\noindent {\bf #1} }


\def \hb {\hfill\break}

\def\lessapprox{\hbox{$\lower1pt\hbox{$<$}\above-1pt\raise1pt\hbox{$\sim$}$}}
\def\greaterapprox{\hbox{$\lower1pt\hbox{$>$}\above-1pt\raise1pt\hbox{$\sim$}$}}

\bls{16}
$ $
\vskip 1.8truecm

\title{
\underbar{
ELECTRON-PHONON DRIVEN SPIN FRUSTRATION IN MULTI-BAND
}\hfil\break \underbar{
HUBBARD MODELS: MX CHAINS AND OXIDE SUPERCONDUCTORS}
}
\authors{
J.~Tinka GAMMEL,$^{(a)}$ K.~YONEMITSU,$^{(b)}$
A.~SAXENA,$^{(b)}$ A.R.~Bishop,$^{(b)}$
and H.~R\"oder$^{(c)}$\hb
$^{(a)}$ Code 573, Materials Research Branch,
NCCOSC RDT\&E Division (NRaD),
San Diego, CA 92152-5000, USA\hb
$^{(b)}$ Theoretical Division and Centers for Nonlinear Studies and Materials
Science, Los Alamos\hb
National Laboratory, Los Alamos, New Mexico 87545, USA\hb
$^{(c)}$ Physikalishes Instit\"ut, Universit\"at Bayreuth,
W-8580 Bayreuth, Germany
}
\header{ABSTRACT}

We discuss the consequences of both electron-phonon and electron-electron
couplings in 1D and 2D multi-band (Peierls-Hubbard) models.  After briefly
discussing various analytic limits, we focus on (Hartree-Fock and exact)
numerical studies in the intermediate regime for both couplings, where
unusual spin-Peierls as well as long-period, frustrated ground states
are found.  Doping into such phases or near the phase boundaries can lead
to further interesting phenomena such as separation of spin and charge,
a dopant-induced phase transition of the global (parent) phase, or
real-space (``bipolaronic'') pairing.  We discuss possible experimentally
observable consequences of this rich phase diagram for halogen-bridged,
transition metal, linear chain complexes (MX chains) in 1D and the oxide
superconductors in 2D.


\header{INTRODUCTION}

The MX chain {\it class} of materials [1]
and oxide high-temperature superconductors (HTS)
are striking similar from the perspective of strong competitions for
broken-symmetry ground states ---
bond-order-waves (BOW), charge- (CDW) or
spin- (SDW) density-waves or antiferromagnetism (AF),
or both bond- and spin- distortions
(spin-Peierls, SP) --- and the properties of
doping- and photo-induced local defect states (kinks,
polarons, bipolarons, excitons, breathers, {\it etc.})
within the same wide variety of novel ground states.
It is becoming increasingly apparent that to correctly
model these materials, not only must one
take into account both electron-electron (e-e)
and electron-phonon (e-p) interactions, but also that
multi-band and/or multi-orbital effects are
critically important.

Experimentally, the remarkable {\it tunability} of
the MX class, along with the ease of synthesis of single crystals, is a
major advantage allowing systematic investigation between small and large
polaron regimes, the influence of dimensionality, and the competition
with non-adiabatic and impurity/localization effects.
Intrinsic defects (self-trapped local defect states or bags) in both
HTS and MX chain materials are thought to be polaronic in nature,
and can be created
via doping or photoexcitation.
Electron-hole asymmetry, which has been proposed as a driving
mechanism for superconductivity (SC),
has been experimentally observed in MX compounds [2].
Recent evidence in PtBr suggests bipolaron dissociation and
polaronic trapping as temperature is varied [3].
Moreover, the model described below manifests long-period (LP)
(``superlattice") phases [4],
possibly observed in recent experiments on MX
compounds, and may be related to twinning or real space pairing in the
cuprate HTS.  Importantly, SC is believed to
occur at the regions of ``melting" between various broken symmetry
ground states and correlated metals.
Thus we anticipate the possibility of
SC, or one dimensional (1D) precursors of SC,
in MX materials when we are near metal-insulator
boundaries, where bipolarons are extended and overlapping, competing
with a correlated metal state.  Here also the tunability of the MX class
can be utilized, leading to our current strategy of studying PtI and
Pd$_x$Ni$_{1-x}$Br to tune into weak CDW and SDW regimes.
Recent unusual results on PtI in high magnetic fields [5]
may indicate such a 1D precursor.

Theoretically,
MX materials are clearly a 1D template [6]
for the same many-body methods and
parameter determinations one must employ in 2D cuprate and 3D bismuthate
HTS models when non-bonding orbitals are neglected.  Particularly
all are hybridized, multi-band materials.
The appearance of superlattice
CDW ordering in MX materials [4]
may be similar to observations in many HTS
materials.  In other words including a microscopic driving force for
finite scale ``twinning" textures and integrating electronic degrees of
freedom yield an effective anharmonic lattice dynamics.  Indeed, in the
model described below the superlattices arise from ordering of
``bipolaronic" defects with respect to the period-4 CDW, corresponding
to a ``melting" of the broken symmetry state
({\it cf.} Ba$_{1-x}$Pb$_x$BiO$_3$).  Polaron
pairing into bipolarons is a common ingredient of many current e-e and
e-p theories of HTS materials.
Further,
both MX chains and (with M=Cu, X=O) CuO$_2$ planes (and chains),
can be described by essentially the same multi-band, tight-binding
extended Peierls-Hubbard Hamiltonian (PHH),
described briefly below, although a ${3/4}$-filled,
2-band (2B) 1D model for the MX
compounds or a CuO chain, and a
${5/6}$-filled, 3-band 2D model for the CuO planes.
This same model can be considered
a 3/4-filled analog of the the organic conductor polyacetylene,
model charge-transfer salts such as TTF-TCNQ, or be used to
investigate neutral-ionic transitions [7].
Mathematically, one can consider the two orbitals to be on the same site,
and thus this Hamiltonian is also related to the Kondo Hamiltonian
used to describe heavy fermion materials.

\header{MULTI-BAND MODEL}

Our model 1D, 2B, PHH
focusing on the d$_{z^2}$ and p$_z$ orbitals
and including only the nearest neighbor interactions
is [8,9]
$$\eqalign{
H =&
\sum_{l,\sigma} \biggl\{\bigl(-t_0 +  \alpha \delta_l\bigr)
                \bigl(c^\dagger_{l,\sigma} c_{l+1,\sigma}
                     +c^\dagger_{l+1,\sigma} c_{l,\sigma}\bigr)~
+~\bigl[\epsilon_l-\beta_l (\delta_l+\delta_{l-1})\bigr]
               c^{\dagger}_{l,\sigma} c_{l,\sigma} \biggr\}
\cr
&+\sum_{l} \biggl\{
U_l n_{l\uparrow} n_{l\downarrow}
+\, V n_{l} n_{l+1}
+\, {1\over 2} K (\delta_l-a_0)^2 ~,
}\eqno{(E1)}$$
where $(\epsilon,\beta,U)_{l}$=$(\epsilon,\beta,U)_{M,X}$
and $\epsilon_M$$-$$\epsilon_X$=$2e_0$.
For a more detailed description of the parameters,
the methods of solution, and general properties
of this Hamiltonian, see Ref.s [8] and [9].
The results here were obtained by exact diagonalization (ED) of
the many-body PHH on small systems.

Tuning $e_0$, $t_0$, $U_M$, and $U_X$ is essentially a 1D
analog of the
theoretical discussions in 2D which focus on the $p$-$d$
hybridization in HTS materials.
There, as here, these parameters determine the stoichiometric
ground state broken symmetry order, and the nature of electron
or hole doping into those ground states. Indeed,
the similarity is even stronger because the nominal band filling
is essentially the same in both MX and HTS materials --
3/4-filling of 2 bands and 5/6-filling of 3 bands, respectively.
Furthermore, both strongly AF and CDW stoichiometric
compounds exist for both HTS and MX. We stress
the MX {\it class} has the advantage of essentially continuous
tunability between these extremes.
Examining trends in related materials often yields
insights that a single material focus might easily miss.
Finally, we note that
competition between e-p and e-e
coupling is especially important for for doping states
and can induce strong {\it local}
LD even in a strongly magnetic background [9,10].

\header{ANALYTIC LIMITING CASES}

In the LD CDW case typical of many MX chains, a first
approximation is to approach the material as decoupled
X=M=X trimers and M monomers. Indeed, PtCl chain
absorptions are very close to the monomers and
trimers in solution. Conversely, when e-e interactions
are dominant and the splitting between the bands is large
(1B limit), one expects a SP phase
in 1D [11], with an effective AF coupling
between M sites
given by $J_{MM}=t_{MM}^2/U_M$, where
$t_{MM}$ is the M-M hopping in the effective 1B model.
We discuss here the $t_0$=0 limit, which captures the essential
features of the material in both these limits.
Another analytically tractable limit ($U$=0) was discussed in detail
in Ref.~[8].

The distortion and energies of the
period-4 (P4) phases for $t_0$=$\alpha$=0, listed in Table 1,
are: two undistorted configurations with unpaired electrons
on the M (MAF) or X (XAF) sites, one phase with both LD
and unpaired spins (MIX) and 2 diamagnetic phases with the M (BOW)
or X (CDW) sublattice distorted and a charge-density-wave
on the opposite sublattice.
The phase diagram for parameters we expect to be relevant to
MI or NiX materials, is shown in Fig.~1.
The phase diagram is more complex for
$U_X$,$\beta_X$$\greaterapprox$$U_M$,$\beta_M$,
where the hybridization-driven competition is most effective.
The BOW (XAF) phase is only found for $|\beta_X/\beta_M|$
($U_X/U_M$) larger than the case shown
(though at $t_0/e_0\sim 2$, or smaller with $\alpha$$>$0,
a BOW phase is found numerically
near the MAF phase boundary for these parameters).

\topinsert
\vglue 2.5truein \bls{12} \noindent
Fig.~1.
The $t_0$=0 phase diagram
for $\beta_X/\beta_M$=$U_X/U_M$=1.
The dashed line is the slice used
for Fig.~4 and + the point in Fig.~3.
Note changing $P$ translates into moving along
a line through the origin.
$$\vbox{
\noindent Table 1.
The period-4 phases, occupancies,
distortions, energies, and effective $J$ scaling
in the $t_0$$\rightarrow$0 limit.
Here $u_{m,x}$=$U_{M,X}/e_0$, $b_{m,x}$=$(\beta_{M,X})^2/(Ke_0)$, and
$ \delta(n)$=$ d_X (\cos {n\pi\over2}$$-$$\sin {n\pi\over2} )
          $$-$$d_M (\cos {n\pi\over2}$$+$$\sin {n\pi\over2} ) $
defines X(M) sublattice distortion order parameters $d_X$($d_M$)
with the first bond (X-M) being short.
(constant or $(-1)^n$ terms just renormalize $t_0$ and $e_0$).
For non-period-4 phases see Ref.~[4].
For $t_0$$\ne$0, the unpaired spins are
antiferromagneticly correlated.
\vskip6truept
\settabs \+~~~~~~~~~~~~~~MIX  &~~~~~~1=1$-$2$\cdot$$\cdot$$\cdot$2$-$
     &~~~~~~$\beta_M/(2K)$
     &~~~~~~$\beta_X/(2K)$
     &~~~~~~$u_x/2$$-$$b_x/4$$+$$u_m/2$$-$$b_m/4$
     &~~~~~~$J_{MX}$$\propto$$t_0^2$\cr
\hrule
\smallskip
\hrule
\smallskip
\+~~~~~~~~phase &$X\,M\,X\,M~$
     &~~~$d_M$
     &~~~$d_X$
     &~~~~~~$2E_T/(e_0N)$
     &~~~$J^{\rm eff}$\cr
\smallskip
\hrule
\smallskip
\+~~~~~~~~CDW
&$\!\uparrow\!\!\downarrow$=$\bullet$=$\!\uparrow\!\!\downarrow$$-$$\!\uparrow\!\!\downarrow$$-$
     &~$\beta_M/K$
     &~~~~0
     &~~~$u_x$$-$$1$$+$$u_m/2$$-$$b_m$
     &~~~~--\cr
\+~~~~~~~~BOW
&$\bullet$=$\!\uparrow\!\!\downarrow$$-$$\!\uparrow\!\!\downarrow$$-$$\!\uparrow\!\!\downarrow$=
     &~~~~0
     &~$\beta_X/K$
     &~~~$u_x/2$$-$$b_x$$+$$u_m$$+$$1$
     &~~~~--\cr
\+~~~~~~~~MIX
&$\uparrow$=$\downarrow$$-$$\!\uparrow\!\!\downarrow$$\cdot$$\cdot$$\cdot$$\!\uparrow\!\!\downarrow$$-$
     &$\beta_M/(2K)$
     &$\beta_X/(2K)$
     &$u_x/2$$-$$b_x/4$$+$$u_m/2$$-$$b_m/4$
     &$J_{MX}$$\propto$$t_0^2$\cr
\+~~~~~~~~MAF
&$\!\uparrow\!\!\downarrow$$-$$\uparrow$$-$$\!\uparrow\!\!\downarrow$$-$$\downarrow$$-$
     &~~~~0
     &~~~~0
     &~~~~~~~~$u_x$$-$$1$
     &$J_{MM}$$\propto$$t_0^4$\cr
\+~~~~~~~~XAF
&$\uparrow$$-$$\!\uparrow\!\!\downarrow$$-$$\downarrow$$-$$\!\uparrow\!\!\downarrow$$-$
     &~~~~0
     &~~~~0
     &~~~~~~~~$u_m$$+$$1$
     &$J_{XX}$$\propto$$t_0^4$\cr
\smallskip
\hrule
\smallskip
\hrule
}$$
\endinsert
\bls{16}

\header{SPIN-PEIERLS AND FRUSTRATED PHASES}

When $t_0$$\equiv$0, all spin excitations are isoenergetic.
For $t_0$$\ne$0, AF correlations develop and
one can treat the problem in terms of an
effective spin model [12].
Without coupling between the
two bands, the lower, X-like (for $e_0$$>$0) band
is full and non-magnetic, while the upper (M-like)
band is 1/2-full with one electron per M
site (when e-e repulsion
is dominant). The on-site e-p coupling $\beta$ leads to a
splitting of the upper band which competes with
an AF ordering of the spins
caused by the effective AF coupling ($J_{MM}$)
(as in the effective 1B case).
When hybridization between the two bands is allowed,
the lower band is not completely full, and now there is
an effective AF coupling between neighboring halide sites
($J_{XX}$, dominant when $U_X$ is dominant)
as well as M and X sites ($J_{MX}$), not present in 1B models.
In fact,
when the splitting due to the e-p coupling $\beta$
is on the order of $U$ and $e_0$, the AF state with neighboring
M-X pairs singly occupied can become the ground state, as shown
in Fig.~1.
Note this implies, in contrast to the 1B case, that
the combination of e-e and e-p coupling in the 2B model drives,
in addition to the non-magnetic CDW and BOW phases,
three (competing) SP phases: one on the
X-sublattice (XAF), one on the M-sublattice (MAF),
and one involving MX pairs (MIX).
Since $J_{MM}$, $J_{XX}$, and $J_{MX}$ are all AF couplings
($J>0$), they obviously cannot all be simultaneously satisfied
and the system is frustrated,  as shown in Fig.~2.
It is straightforward to derive
from fourth-order perturbation theory
in the large $U$ limit an
effective $t$, $J_{MX}$, $J_{MM}$, $J_{XX}$ model,
though the expressions for the $J$'s are cumbersome and
phase dependent [12].
The resultant $t_0$ dependencies of the
$J$'s are listed in Table 1.
When $t_0$ is small and one is in a regime where
only one of the $J$'s
is important, one can numerically check this estimate by
comparing the energies of the singlet and triplet ground states
(at fixed LD).
Fig.~3 shows the $t_0$ dependence is correctly predicted.
Note the CDW phase has an entirely e-p driven AF component.

\topinsert \vglue 2.5truein \bls{12} \noindent
Fig.~2 (left).
Schematic energy level diagram in the strongly correlated
limit showing frustration of the effective antiferromagnetic couplings
caused by MX hybridization.
\vskip 6truept \noindent
Fig.~3 (right).
The $t_0$ dependence of the difference in singlet and triplet
energies for the CDW, MIX, and MAF phases
at the + in Fig.~1.
For small $t_0$, this corresponds to excitations
of the effective spin-Hamiltonian $H=4JS_i S_{i+j}$
with energy $4J_{MM}$ in the MAF phase and
$4J_{MX}$ in the MIX phase,
whereas in the CDW phase this reflects the Peierls gap.
\vglue 2.5truein \noindent
Fig.~4.
(a) Total energy and
(b) lattice distortion amplitude as a function of {\it U}$_M$
for CDW ($-$), BOW (-~-), SP ($\cdot\cdot\cdot$) SDW ($-\cdot$),
and LP ($-$ $-$) phases.
Parameters consistent with CuO were used.
\endinsert \bls{16}

In Fig.~4 we show
the total energy and average lattice distortion for
parameters similar to those used in the
2D model for CuO planes [10]
(in hole notation, $e_0$=$t_0/2$$-$$U_M/4$, $\beta$=0,
$U_X/t_0$=3, $V/t_0$=1, $\alpha^2/Kt_0$=2, and the M-M distance
was held constant) with analogous results.
For small $U_M$, the ground state is a CDW, as expected [8,9].
As $U_M$ increases, the X no longer symmetrically distort
(BOW in notation of Ref.~[10]), then a LP phase develops,
and finally the SDW phase expected at large $U_M$ sets in.
In contrast to the 2D case, the SP
phase is not seen, though a metastable SP
regime is found.
As is clearly seen from the figure, the ``melting"
of the CDW phase takes place through a LP
intermediary [13].
SC is believed to
occur at the regions of ``melting" between various broken symmetry
ground states and correlated metals.
While SC is not expected in the 1D materials,
unusual results on PtI in high magnetic fields [5]
may be a 1D precursor.

Upon doping, the phase diagram becomes even richer.
As has also been seen in the 2D version of the PHH [10],
doping into, $e.g.$, the SDW phase can lead to polaronic
defects with local CDW character, and $vice$-$versa$.
An example of this for our 1D model was reported in Ref.~[9]
for Ni parameters.
Similar results have recently been
reported in a 2D model for the HTS [10].
Further, near the phase boundary, we have also found
regions where doping into the SDW ground state can cause
a CDW defect {\it in a CDW background}  (and $vice$-$versa$)
-- $i.e$, the
dopant has altered not only it's local environment, but
the background phase as well.
While clearly finite-size effects will be important
for this energy balance, one can easily find parameter
regions where the ``finite size" is as large as
typical correlation lengths in the real materials.
Additionally, in the  ``zero-hopping" limit where the
many-body PHH is analytically tractable,
one can show that such phase reversal upon doping can occur
even in the infinite-system limit.
Indeed, the narrow CDW phase seen near 1/8th doping
in the La based HTS compounds
may be related to this phenomenon.

The crossover from P4 CDW to P4 MAF may also be
accompanied by LP superlattice phases
[14]. Such superlattice phases have recently
been found in a 2D, 3B model of HTS CuO$_2$ layers, including
{\it intersite} e-p coupling.
We anticipate the same behavior in
our MX model when intersite e-p coupling is included.
Such superlattices may be viewed as an ordered array of
discommensuration defects with respect to a nearly stable commensurate
order ($e.g.$, P4). In view of the effective $J$'s discussed
above, it may be natural to model such states in terms of ANNNI-like
models, where nearest and longer range couplings
compete, leading to frustration and associated complexity
phenomena -- multitimescale relaxation, hysteresis, metastability,
$etc$. In the context of the MX class, it will be
particularly interesting to investigate materials in, or near,
this crossover regime -- $e.g.$, PtI -- and to further control
the crossover with pressure,
magnetic field [5],
doping, impurities, $etc$. Indeed {\it doping} into this
complex regime should be highly sensitive to the softness
and competitions of the phases: This may well be an excellent
regime to study pairing tendencies and metallization.

\header{CONCLUSIONS}

We stress that the 1D, 2B, PHH, while simple to write down,
is representative of a very large variety of low-D
electronic materials. This variety is mirrored in the model's
richness, especially in terms of doping near phase boundaries
where novel pairing mechanisms are found
(inclusion of e-p in a 2D 3B model leads to coexistence
of CDW and SDW phases -- ``charge bags" --
as e-p interactions are important only in the
neighborhood of defects [10]),
or a dopant-induced transition [15]
of the $global$ phase may exist,
besides the usual plethora of doping and photoinduced
non-linear excitations.
The MX {\it class} of compounds
are uniquely important as a testing ground for many-body modeling
and materials design strategy in strongly correlated,
low-D materials (in particular the oxide HTS).
Apart from pure and mixed-halide MX materials we are also
beginning to explore mixed-metal (M$_xM'_{1-x}$X)
and bimetallic (MMX) systems, as well as effects of magnetic
fields (especially on the weak CDW/SDW ground state materials).
We feel that experimental investigations of the pressure
dependence and
high (magnetic) field behavior of pure and doped materials
in the LD/AF cross-over regime, such as
PtI, NiBr, or their mixed-metal or -halide analogs,
will continue to yield interesting insights into the nature
of multiband effects and the competition between e-e
and e-p interactions.


\underbar{\it Acknowledgements}.
We acknowledge important discussions with X.Z.~Huang,
L.A.~Worl, R.~Donohoe, and B.I.~Swanson,
J.~Shi, among others,
and thank E.Y.~Loh, Jr.~for help in writing the ED code.
ARB, KY, and AS were supported by the US DOE,
HR by the Deutsche Forschungsgemeinschaft through SFB 213,
and JTG by a NRC-NRaD Research Associateship
through a grant from ONR.
Supercomputer access was through the ACL
at LANL. HR and JTG are grateful
to the hospitality of LANL, where this work was begun.

\header{REFERENCES}

\item{1}
The structure and chemistry of MX
chains, [ML$_2$][ML$_2$X$_2$]$\cdot$Y$_4$
(M: Pt,Pd,Ni; X: Cl,Br,I; L,Y: various ligands, counterions),
has been reviewed by H.J.~Keller, in J.S.~Miller (ed.),
{\it Extended Linear Chain Compounds},
Vol.~1, Plenum, New York, 1982, p.~357.
For a sampling of the physics literature see, {\it e.g},
Proc.~of the Int.~Conf.~on
the Sci.~and Tech.~of Synth.~Metals,
ICSM '90, T\"ubingen, Germany, Sept.~2-7, 1990
[{\it Synth.~Metals} {\bf 41-43}, (1991)], or
Ref.s [8] and [9] and references therein.
\item{2}
J.T.~Gammel {\it et al.},
{\it Phys.~Rev.~B}, {\bf 42} (1990) 10566.
\item{3}
R.J.~Donohoe {\it et al., J.~Phys.~C}, (1992), in press.
\item{4}
I.~Batisti\'c, J.T.~Gammel and A.R.~Bishop,
{\it Phys.~Rev.~B}, {\bf 44} (1991) 13228.
\item{5}
M.~Haruki and P.~Wachter,
{\it Phys.~Rev.~B}, {\bf 43} (1991) 6273.
\item{6}
D.~Baeriswyl and A.R.~Bishop,
{\it Physica Scripta},  {\bf T19} (1987) 239;
S.D.~Conradson {\it et al.}, {\it Solid State Commun.}, {\bf 65} (1988) 723;
J.T.~Gammel {\it et al.},
{\it Physica B}, {\bf 163} (1990) 458.
\item{7}
A.~Painelli and A.~Girlando,
{\it Synth.~Metals} {\bf 29} (1989) F181.
\item{8}
J.T.~Gammel {\it et al.},
{\it Phys.~Rev.~B}, {\bf 45} (1992) 6408.
\item{9}
S.M.~Weber-Milbrodt {\it et al.},
{\it Phys.~Rev.~B}, {\bf 45} (1992) 6435.
\item{10}
K.~Yonemitsu {\it et al.}, to be published in the
Proc.~of the Int.~Conf.~on Lattice Effects in High T$_c$ Superconductors,
Sante Fe, New Mexico, Jan.~13-15, 1992 (World Scientific).
\item{11} 
J.E.~Hirsch, {\it Phys.~Rev.~Lett.}, {\bf 51} (1983) 296.
\item{12}
J.~Shi, R.~Bruinsma, and A.R.~Bishop, preprint 1992.
\item{13}
There are many metastable LP phases, and the LP energies
should be treated
as upper bounds. From Fig.~4 there is clearly a region where
the LP phase is favored over any period-4 phase.
This intermediate LP region is seen in ED,
MFA, and the 2D model.
It is not an artifact.
\item{14}
Intermediate phases can be driven by the competition
between $t_0$, $\beta$, $\alpha$, long-range Coulomb fields,
and/or the effective short range $J$'s. Such intermediate phases
can be P4, as expected from commensurability effects, or LP.
Note that the large-$\beta$ LP phase
discussed in Ref.~[4], though related,
is somewhat different than the frustration driven, magnetic LP phase
discussed here or, in 2D, in Ref.~[10]
(for large $\alpha$).
Similar to the ANNNI model,
no explicit long range terms are needed as
frustration drives effective long range interactions,
\item{15}
J.~Reichl, unpublished and S.~Marianer, unpublished.

\end